\documentclass[aip,jap]{revtex4-1}

\usepackage{graphicx,amsmath,amssymb,subfig}

\begin{document}
\title{The role of electron-electron scattering in spin transport} 
\author{Akashdeep Kamra}\email{akashk@iitk.ac.in}
\author{Bahniman Ghosh}
\affiliation{ Department of Electrical Engineering, IIT - Kanpur, Kanpur - 208016, India}

\date{\today}

\begin{abstract}
We investigate spin transport in quasi 2DEG formed by \textrm{III-V} semiconductor heterojunctions using the Monte Carlo method. The results obtained with and without electron-electron scattering are compared and appreciable difference between the two is found. The electron-electron scattering leads to suppression of Dyakonov-Perel mechanism (DP) and enhancement of Elliott-Yafet mechanism (EY). Finally, spin transport in InSb and GaAs heterostructures is investigated considering both DP and EY mechanisms. While DP mechanism dominates spin decoherence in GaAs, EY mechanism is found to dominate in high mobility InSb. Our simulations predict a lower spin relaxation/decoherence rate in wide gap semiconductors which is desirable for spin transport.  
\end{abstract}


\maketitle

\section{Introduction}
In the last couple of decades, there has been an increasing effort towards exploring the possibilities of exploiting the spin degree of freedom of electrons in practical devices. Hence, a new field called Spintronics has emerged \cite{fabian,bandyopadhyay,zutic,wu}. While metal based structures have seen some success in the form of Magnetic Tunnel Junctions, semiconductors still remain to make an impression. One important question is what type of semiconductors should be preferred for use in spintronics. So far there has not been any definitive answer to this issue.

Realization of devices based on spin of the electron broadly involves a successful implementation of three processes - spin injection, spin transport and spin detection. It is the second process - spin transport - that we focus on in the present work. The phenomenon of spin decoherence is the major ingredient to any study of spin transport. The spin relaxation mechanisms that are pertinent to an $n$ type heterostructure are Dyakonov-Perel mechanism \cite{dp} (DP) and Elliott-Yafet mechanism \cite{ey} (EY). The former has been believed to dominate the latter in high mobility systems. 

We pause to discuss the discrimination that we make between the terms {\it spin relaxation} and {\it spin decoherence}. The term {\it spin relaxation} is appropriate for an initially polarized homogeneous system which looses its spin polarization with time so that the system relaxes to its equilibrium state. On the other hand, in spin transport, a steady state is achieved in which electrons with identical spin polarization are injected at one end of the device. The injected electrons can be said to be coherent at the injection boundary in the sense of having a common spin polarization direction. As the electrons traverse the device, their polarizations spread about an average polarization direction leading to decoherence. Now this average polarization direction can rotate several radians before it loses its meaning due to substantial decoherence. The length over which this substantial decoherence takes place is referred to as the {\it spin relaxation length} in the literature. We will continue to refer to it by its traditional name although we understand that it is actually the spin decoherence length according to our interpretation. We also continue referring to the DP and EY mechanisms as spin relaxation mechanisms although we understand that their role in the context of spin transport is to cause decoherence.

The Monte Carlo method \cite{jacoboni,lugli} has been a very popular technique for studying electronic transport in all its generality. However, most electronic transport studies based on this method ignore electron-electron scattering \cite{warmenbol,kocevar,yokoyama,chattopadhyay}. This is reasonable as the conventional charge current is not affected in a major way by electron-electron scattering as the latter does not directly contribute to momentum relaxation. The electron-electron scattering merely leads to a redistribution in $k$ space which is not of such a high importance in most conventional electronic systems. This tradition of not including electron-electron scattering has been carried over in one of the early works on Monte Carlo method for spin transport \cite{privman}, and in several subsequent works \cite{kamra,tierney2,saikin}. Given the nature of DP and EY spin relaxation mechanisms, electron-electron scattering is bound to be of utter importance. Glazov and Ivchenko have also evinced the importance of electron-electron scattering in calculating spin relaxation time in a spatially homogeneous system \cite{glazov,glazov2}. Wu and co-workers have consistently included electron-electron scattering in their investigations of spin transport using kinetic spin Bloch equations \cite{wu1,wu2,wu3,wu4}. 

We study spin transport in quasi 2DEG using the Monte Carlo approach. The simulation results have been obtained with and without the inclusion of electron-electron scattering for a direct comparison. As anticipated, the electron-electron scattering has been found to play an important role. It leads to a faster equilibration of the electrons and suppression of the DP spin relaxation mechanism. In the light of the importance of electron-electron scattering, one could have expected a diminished strength of DP mechanism and an enhanced role of EY mechanism \cite{songkim}. Hence, we go one step further to include EY mechanism in our simulation routines and obtain results for InSb, which has been treated as a representative of narrow gap semiconductors, and GaAs, which represents wide gap semiconductors. While EY mechanism does not play any role in GaAs, it dominates the spin decoherence in a high mobility InSb heterostructure. Our simulation results establish EY mechanism as the dominant spin decoherence mechanism in narrow gap semiconductors, and indicate a preference towards wide gap semiconductors as the material of choice for spintronics, owing to a lower spin decoherence/relaxation.

\section{Model}
A comprehensive review of the Monte Carlo method and its extension to inclusion of spin degree of freedom has been presented elsewhere \cite{jacoboni,lugli,privman}. We restrict our discussion to enlisting the essential features of our model. The dynamics has been assumed to be confined to the lowest subband. The scattering mechanisms included in all the simulations are acoustic phonon scattering, optical phonon scattering and ionized impurity scattering. The ionized impurity scattering has been treated with Brooks and Herring approach \cite{bh}. In addition, the simulations account for electron-electron scattering wherever mentioned. The band structure has been assumed to be parabolic for calculation of various scattering rates. This restriction has been imposed by our inability to properly account for electron-electron scattering in the non-parabolic band structure scheme. As regards the boundary condition, the electrons absorbed at source end are reintroduced afresh at drain end, and vice versa.

The electron-electron scattering has been treated using Mosko and Moskova's approach \cite{mosko}, which is a refinement of the previously existing techniques \cite{ferry,brunetti}. However, we introduce one change which is instrumental to our model. When an electron undergoes electron-electron scattering, Mosko and Moskova propose to choose its partner electron randomly from the set of all other electrons. This is understandable on the basis of the plane wave nature of electron wave functions considered \cite{mosko}, which specify no information in coordinate space. However, in our semi-classical model, there is a definite coordinate value associated with every electron. This calls for a modification in the method of selecting the partner electron. Choosing the nearest electron at the time of scattering appears to be the most logical alternative. This recipe is of paramount importance in obtaining an accurate distribution in coordinate space.

The spin dynamics is dominated by two relaxation mechanisms - DP and EY. While DP mechanism is considered in all the simulations, EY mechanism is accounted for wherever mentioned. The electrons are evolved under the influence of Rashba \cite{rashba} and Dresselhaus \cite{dresselhaus} Hamiltonians (Eqs. (\ref{dpham})) to account for the DP mechanism.
\begin{eqnarray}\label{dpham}
H_{R} \ & = & \ \alpha \left(k_{y}\sigma_{x} - k_{x}\sigma_{y} \right) \\ \nonumber
H_{D} \ & = & \ \beta <k_{z}^{2}> \left( k_{y} \sigma_{y} - k_{x} \sigma_{x} \right )
\end{eqnarray}  
In order to account for the EY mechanism, the spin flip rate was assumed to be the same as that in bulk material. In bulk \textrm{III-V} semiconductors, spin flip rate is directly proportional to the scattering rate. Hence, EY mechanism can be accounted for by flipping the spin of the electron with a certain probability during every scattering event, where the probability is given by the proportionality constant between spin flip rate and scattering rate \cite{songkim,pikus} (Eq. (\ref{eyconst})).
\begin{equation}\label{eyconst}
p \ = \ A \ \left(\frac{k_{B} T}{E_{g}}\right)^2 \ \eta^2 \ \left( \frac{1 - \eta/2}{1 - \eta/3}\right)^2
\end{equation}
Here $E_{g}$ is the band gap of the material, and $\eta = \Delta/(\Delta + E_{g})$ with $\Delta$ as the spin-orbit splitting of the valence band. $A$ is a parameter that varies between 2 and 6 depending on the scattering mechanism. It has been chosen as 4 for the purpose of simulation. The formalism stated above has been studied and used for phonon and ionized impurity scattering. Its extension to electron-electron EY scattering is validated by the observation that the latter has been found to be roughly equal in strength to the impurity ion EY scattering \cite{tamborenea}.

Our model, however, has some limitations. It is not capable of dealing with systems with too high a mobility due to two primary reasons. Firstly, at high enough mobilities, scattering from remote ions gains importance which has not been considered in our model. Secondly, when electron-electron scattering is not being considered, the transport tends to become ballistic at high mobilities. The Monte Carlo model is equipped to deal with diffusive transport only. For ballistic transport, one needs a model with Landauer-Buttiker kind of formalism \cite{dattabook}. Hence, the highest mobility that we consider in our simulations is nearly $2~m^2V^{-1}s^{-1}$. 

\section{Results and Discussion}
\subsection{Electron-Electron Scattering}
We choose to investigate the effects of electron-electron scattering by simulating spin transport in an $\mbox{In}_{0.53}\mbox{Ga}_{0.47}\mbox{As}$ based heterostructure. A similar structure has already been studied numerically by Saikin et al. \cite{privman} and experimentally by Nitta and coworkers \cite{nitta}. The thickness of the $\mbox{In}_{0.53}\mbox{Ga}_{0.47}\mbox{As}$ layer is $20 nm$, while the length and width of the device are $1 \mu m$ and $2 \mu m$ respectively. The device dimensions have been chosen in such a way that our simulation routines deal with 20000 electrons at a carrier concentration of $10^{12} cm^{-2}$. The dynamics has been assumed to be confined to the lowest subband calculated at $0.2eV$. An impurity concentration of $4\times10^{11} cm^{-2}$ has been assumed to obtain a mobility close to $1~m^2V^{-1}s^{-1}$ at $77K$. The drain-source voltage ($V_{DS}$) is held constant at $0.05 V$ in all the simulations. The Rashba ($\alpha$) and Dresselhaus ($\beta$) spin-orbit coupling constants have been taken as $0.074~eV. \AA $ and $32~eV. \AA^{3}$ respectively \cite{privman,nitta}. The material constants and other parameters have been adopted from a standard manual on Monte Carlo simulations \cite{damocles}. Finally, a basic time step of $1 fs$ was chosen and electrons were evolved for 20000 steps to reach steady state. Further evolution for 5000 steps was carried out, the data for which was recorded for time averaging.

\begin{figure}[hbt]
\subfloat[][]
{\includegraphics[height=6.3cm,width=8.0cm]{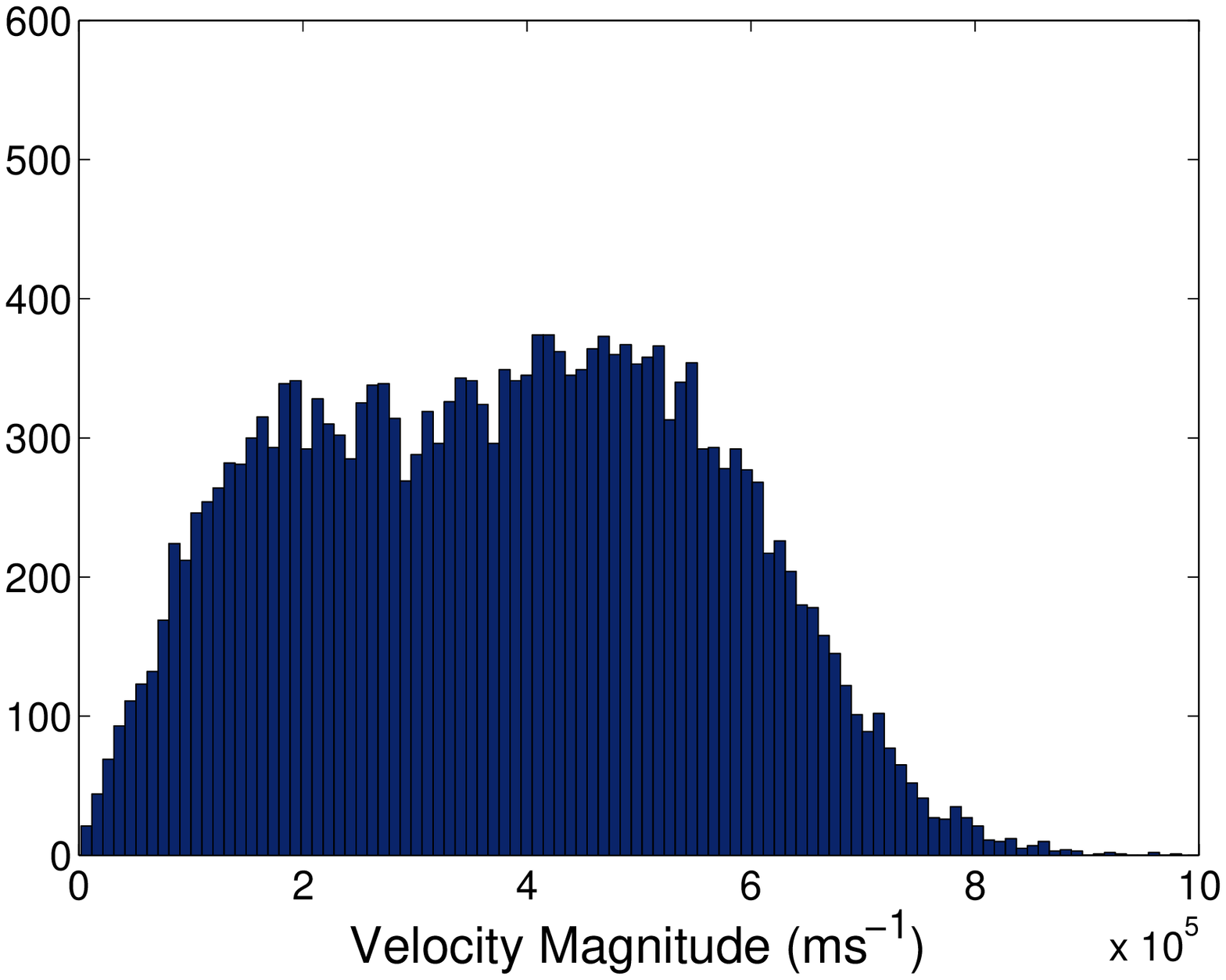}\label{velT77off}}
\quad
\subfloat[][]
{\includegraphics[height=6.3cm,width=8.0cm]{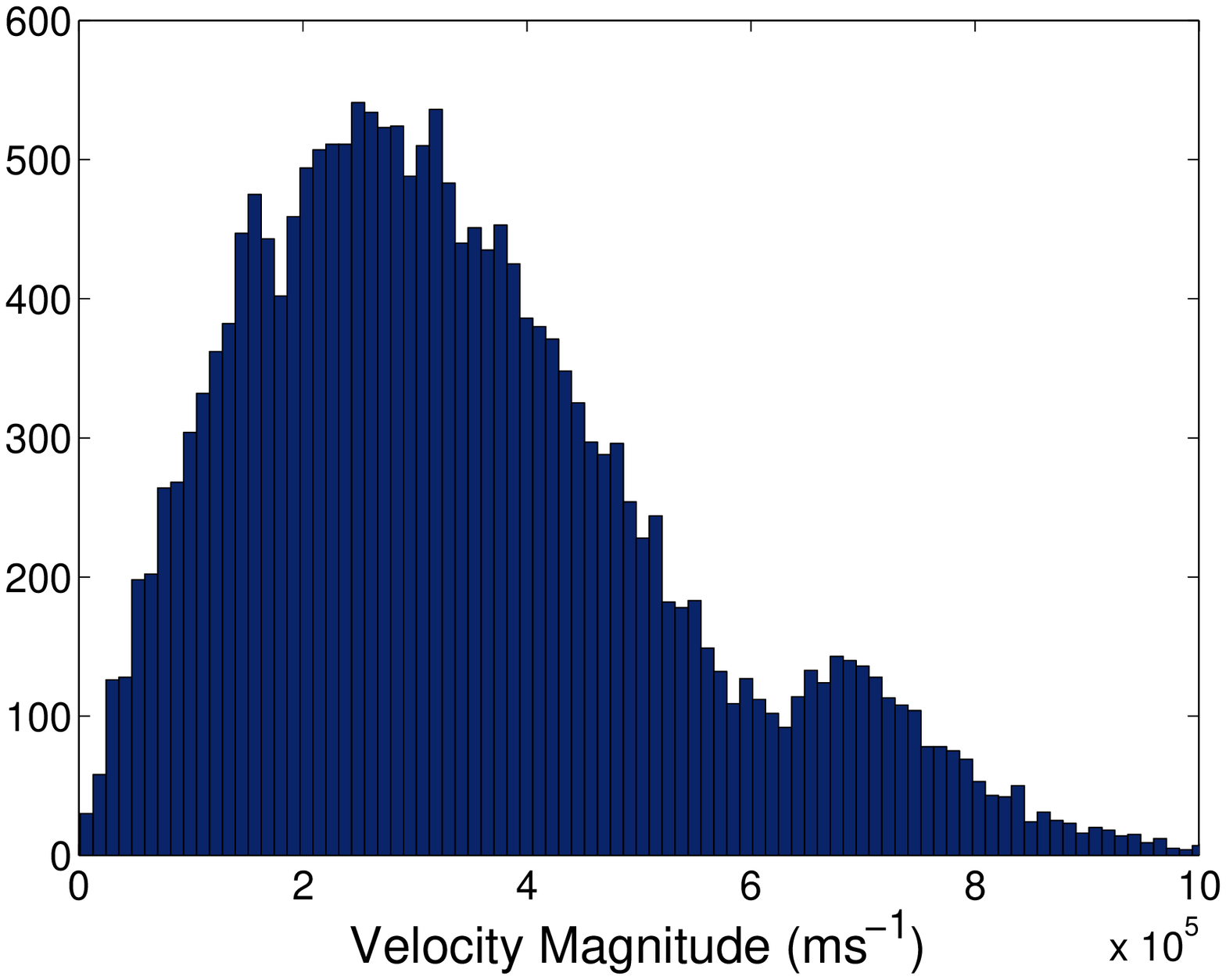}\label{velT77on}}
\caption{Velocity distribution at $T = 77 K$ for the two cases - (a) without (case A) and (b) with (case B) - electron-electron scattering.}
\end{figure}

\begin{figure}[hbt]
\subfloat[][]
{\includegraphics[height=6.3cm,width=8.0cm]{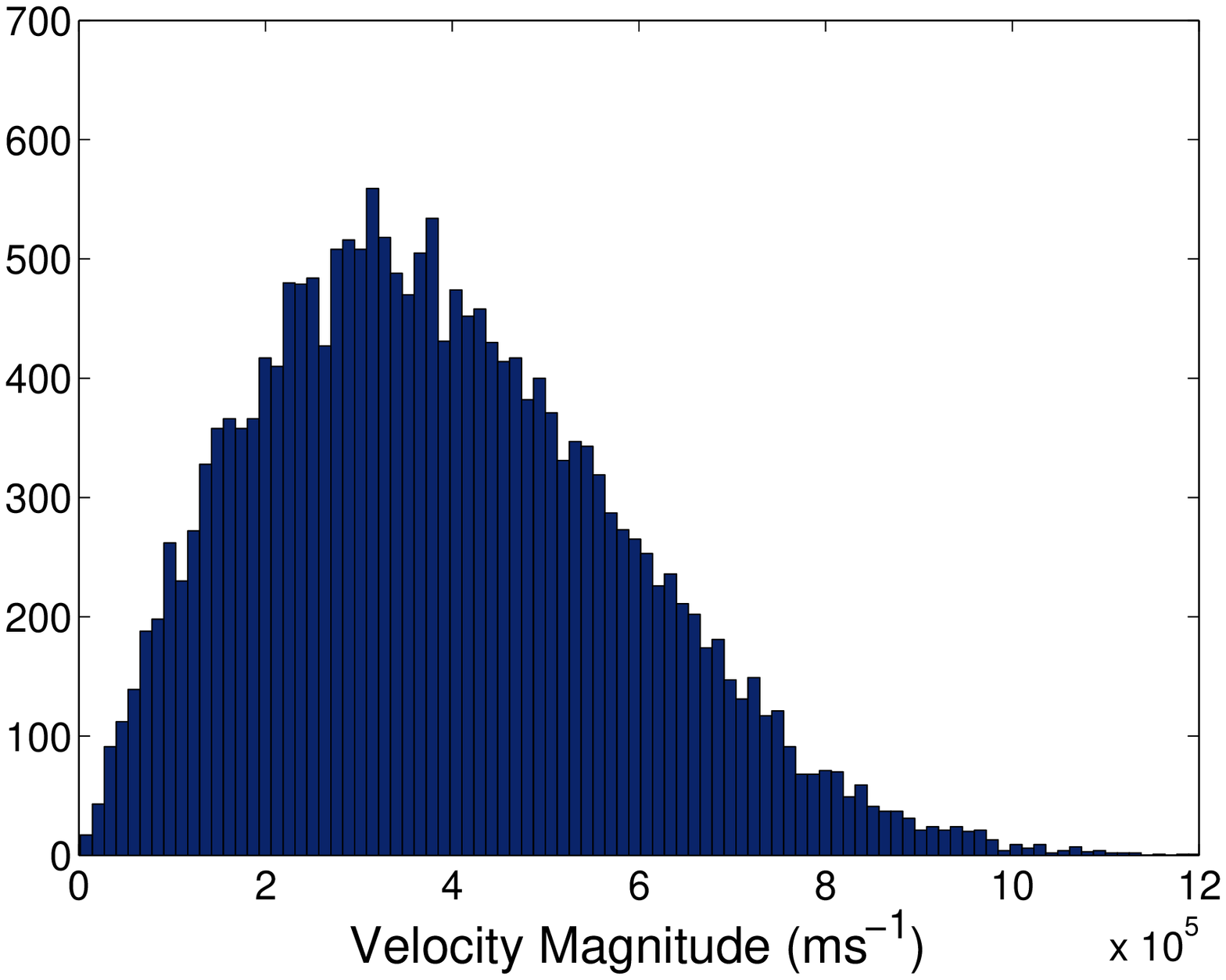}\label{velT300off}}
\quad
\subfloat[][]
{\includegraphics[height=6.3cm,width=8.0cm]{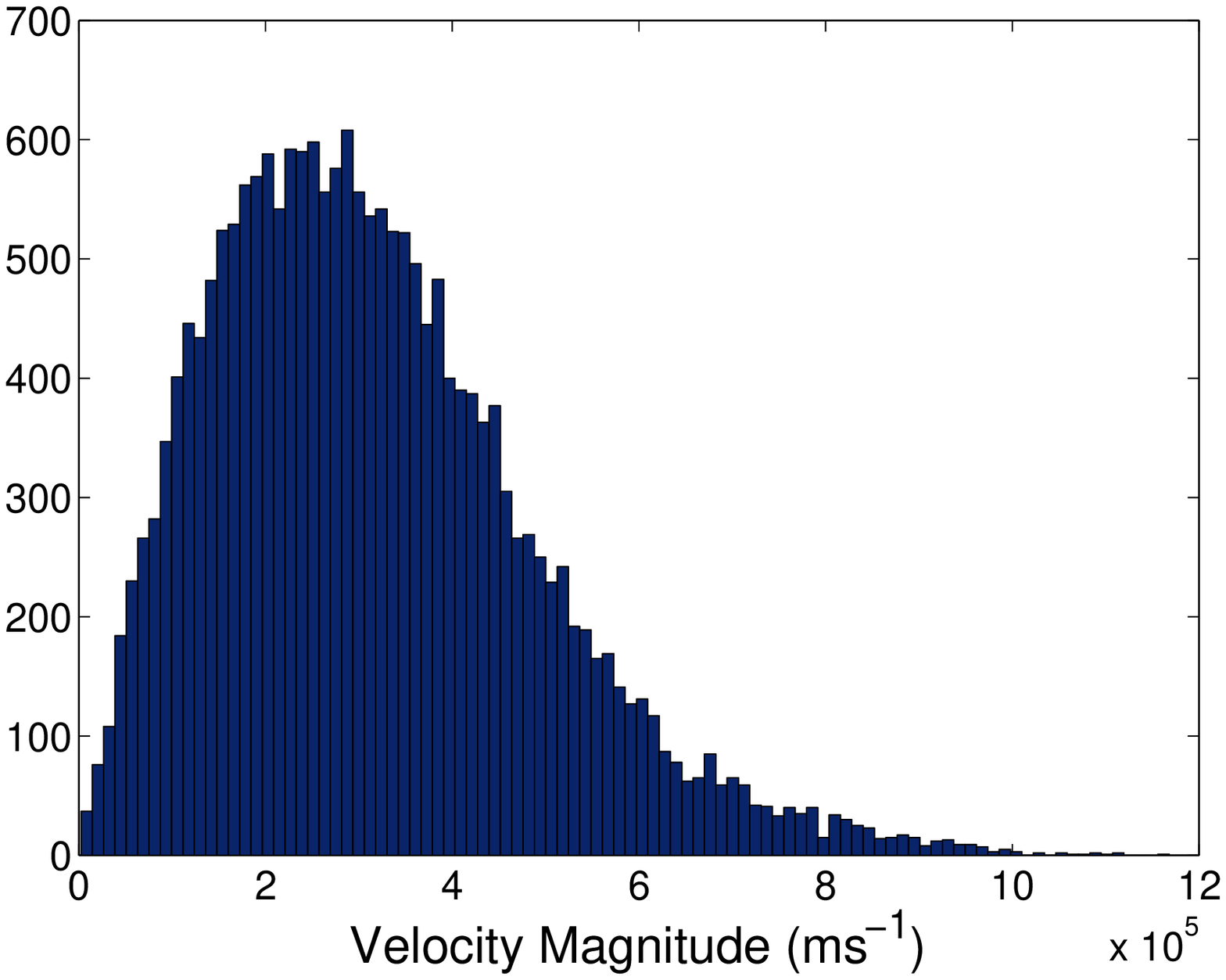}\label{velT300on}}
\caption{Velocity distribution at $T = 300 K$ for the two cases - (a) without and (b) with - electron-electron scattering.}
\end{figure}

The velocity distribution in steady state for the two cases - with and without electron-electron scattering - has been compared in Figs. \ref{velT77off} and \ref{velT77on}. One can clearly see a doubly peaked structure in both the cases. While the velocity distribution is much closer to the expected Maxwell-Boltzmann distribution, with an extra small peak around $7\times10^{5}~ms^{-1}$, when electron-electron scattering is considered (case B), it has two comparable peaks in the other case (case A). The peak towards higher velocity corresponds to the non-equilibrium electron population produced by the electric field. The electrons gain an additional velocity component (having the order of magnitude of $eE\tau_{m}/2m$, where $\tau_{m}$ is the momentum relaxation time and $E$ is the electric field) due to the electric field which, for high mobility systems, gives rise to hot electrons even in steady state. Case B represents a system which is almost at equilibrium and the system is quite far from equilibrium in case A. Hence, one can see that the effect of electron-electron scattering is a faster thermalization of the electrons. Figures \ref{velT300off} and \ref{velT300on} show the same comparison at room temperature. A reversion to unimodal distribution indicates closeness to equilibrium in both the cases. However, case B offers a better fit to the Maxwell-Boltzmann distribution.

\begin{figure}[hbt]
\subfloat[][]
{\includegraphics[height=6.3cm,width=8.0cm]{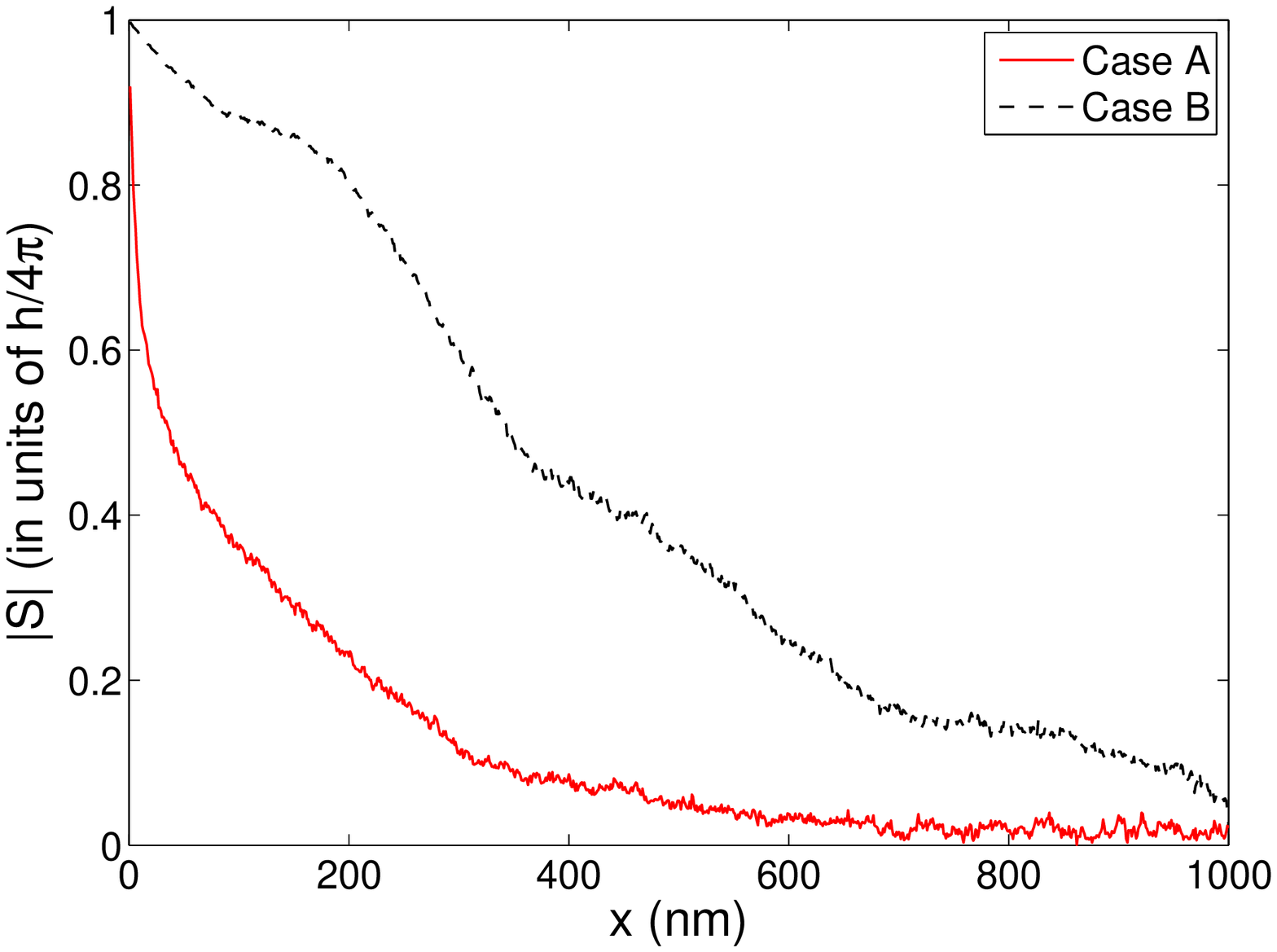}\label{T77meanP}}
\quad
\subfloat[][]
{\includegraphics[height=6.3cm,width=8.0cm]{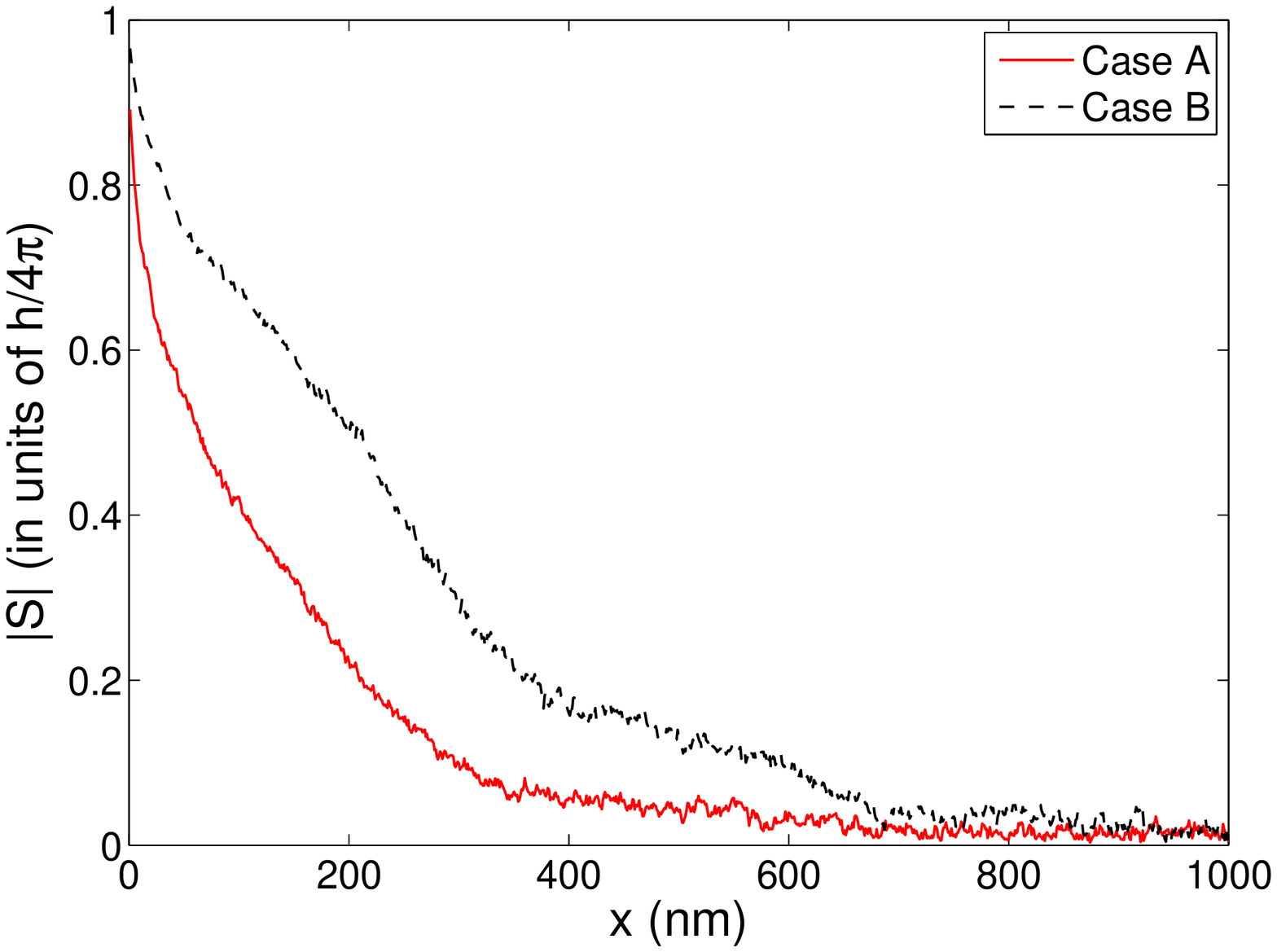}\label{T300meanP}}
\caption{$|S|$ along the channel length at (a) $77K$ and (b) $300K$. Cases A and B stand for exclusion and inclusion of electron-electron scattering in the simulation routines, respectively.}
\end{figure}

We now compare the two cases in terms of their spin space characteristics such as spin relaxation length ($L_{s}$). Figure \ref{T77meanP} depicts $|S|$ along the length of the device at a temperature of $77K$. A striking difference is observed in spin relaxation lengths for the two cases. The electron-electron scattering seems to increase the spin relaxation length drastically. Spin relaxation length for case B is found to be about $4$ times the corresponding quantity in case A. The factor by which the spin relaxation length is increased by electron-electron scattering reduces to about $2$ at room temperature (Fig. \ref{T300meanP}).

The increase in spin relaxation length by electron-electron scattering can be attributed to two reasons. Firstly, electron-electron scattering leads to a narrower distribution in the $k$ space which leads to a reduction in DP spin relaxation as the latter relies on $k$ dependent effective magnetic field for causing spin decoherence. Secondly, what appears to be a more important reason, DP spin relaxation rate is directly proportional to the correlation time $\tau_{c}$ \cite{zutic,songkim} which is reduced by electron-electron scattering. In this context, one needs to make a distinction between the momentum relaxation time $\tau_{m}$ and the momentum redistribution time $\tau_{m}^{'}$, which is roughly equal to $\tau_{c}$. This distinction between the two is a result of the fact that although electron-electron scattering contributes to momentum redistribution, it does not directly lead to momentum relaxation as the momentum remains with the electron body as a whole in case of an electron-electron scattering event. Similar arguments have also been presented by Glazov and Ivichenko \cite{glazov}. The increase in spin relaxation length can be mathematically expressed by the following equations \cite{zutic}.
\begin{equation}
\frac{1}{\tau_{s}}  =  \omega \tau_{m}^{'}
\end{equation}
\begin{equation}\label{ls}
L_{s}^{ee}  =  \sqrt{D\tau_{s}}  = \sqrt{\frac{v_{F}^{2}\tau_{m}}{\omega \tau_{m}^{'}}}  =  \sqrt{\frac{\tau_{m}}{\tau_{m}^{'}}} L_{s} 
\end{equation}
Here $\tau_{s}$ is the spin relaxation time, $\omega$ is a constant, $L_{s}^{ee}$ and $L_{s}$ denote the spin relaxation lengths with and without considering electron-electron scattering, respectively.

The weakening of the effect of electron-electron scattering at room temperature is apparent in the results discussed above. This can be attributed to the relative insensitivity of electron-electron scattering rate to temperature \cite{mosko}, as a result of which, its relative importance begins to decrease with increasing temperature. This can alternately be expressed by stating that the role of electron-electron scattering depends on the ratio $\tau_{m}/\tau_{m}^{'}$ with $\tau_{m}^{'}$ varying slightly with temperature and $\tau_{m}$ decreasing with temperature. In general, this ratio, and hence the role of electron-electron scattering, is larger in systems with a high mobility.

In the absence of electron-electron scattering in the model, calculation of the dominant scattering rate (which is about $10^{12} s^{-1}$) predicts a nearly ballistic transport for a $1\mu m$ long system with high mobility. This can easily be seen by calculating the average distance that an electron travels between consecutive scattering events. Assuming a nominal value of $5\times10^{5} ms^{-1}$ for the electron velocity, this distance comes out to be $0.5 \mu m$. However, electron-electron scattering (which is about $10^{14} s^{-1}$ in the system under consideration) drags the length at which transport becomes ballistic to much below $1\mu m$. This is good news as the Monte Carlo method is appropriate for simulating diffusive transport only. Thus, one can simulate a wider range of systems using Monte Carlo technique, if electron-electron scattering is included in the model. 

Our simulation results demonstrate an important role of electron-electron scattering in spin transport making its inclusion in accurate models inevitable. It affects the velocity distribution by accelerating the equilibration process. More importantly, it leads to a longer spin relaxation length attributed to a diminished DP spin decoherence. The mitigation of DP spin decoherence makes room for a change in the dynamics between various spin relaxation mechanisms. The regions of dominance of DP and EY mechanisms are bound to be affected by the inclusion of electron-electron scattering. We investigate this possibility in the next section.

In their investigation of spin relaxation time in an initially polarized spatially homogeneous electron gas \cite{glazov2}, Glazov and Ivichenko have found a positive role of exchange interaction in suppressing spin relaxation due to DP mechanism. One may be inclined to believe a similar suppression of DP mechanism for the case of spin transport as well. However, the effect of exchange interaction is diminished for the case at hand by the spatial variation of the polarization profile. Since the former is obtained by summing over all the electrons, spatial variation leads to an incoherent summation as opposed to the coherent summation obtained for the homogeneous case. Hence, the exchange interaction in the form used by Glazov and Ivichenko may not be important in spin transport. The exchange interaction has not been considered in the present work.

\subsection{Elliot-Yafet Mechanism}

Most of the literature on spin transport Monte Carlo simulations considers only DP spin relaxation mechanism as it has been believed to be dominant in moderate to high mobility systems. However, in the previous section we have noticed suppression of DP spin decoherence due to electron-electron scattering. Owing to the same reason, one can also anticipate an enhancement in decoherence attributed to EY spin relaxation as it is directly proportional to the scattering rate \cite{songkim}. This motivates us to investigate spin transport taking into account both DP and EY spin relaxation mechanisms and look for any change in the dynamics of spin decoherence dominance. The strength of EY mechanism, however, is strongly dependent on the band gap of the material. Hence, we break our investigation into two parts studying spin transport in heterostructures based on two different materials - InSb, which represents narrow gap semiconductors, and GaAs representing wide gap semiconductors.  

The InSb based heterostructure under investigation is similar to the structure probed in other studies \cite{kamra,gilbertson}. The device parameters and simulation details which are not mentioned here remain the same as for the $\mbox{In}_{0.53}\mbox{Ga}_{0.47}\mbox{As}$ based structure discussed in the preceding subsection. The lowest subband, to which the dynamics has been assumed to be confined, was calculated at an energy of $0.095 eV$. Room temperature was assumed in all the following simulations, at which the system has a mobility of nearly $2~m^2V^{-1}s^{-1}$. The Rashba spin-orbit coupling constant $\alpha$ and $\beta<k_{z}^{2}>$ were taken as $0.15~eV.\AA$ and $0.03~eV.\AA$, respectively \cite{gilbertson}.

As regards the GaAs based heterostructure, a voltage drop of $1 V$ across the $20 nm$ thick GaAs layer was assumed. With this assumption, the lowest subband was calculated at an energy of $0.26 eV$. The mobility of this system at room temperature, which was assumed in the simulation, is about $0.2~m^2V^{-1}s^{-1}$. For the assumed drop in potential across the GaAs layer, the Rashba ($\alpha$) and Dresselhaus ($\beta$) spin-orbit coupling constants are $0.027~eV.\AA$ and $29~eV.\AA^{3}$ respectively \cite{tierney} and $\beta<k_{z}^{2}>$ was calculated as $0.045~eV.\AA$.

\begin{figure}[hbt]
\subfloat[][]
{\includegraphics[height=6.3cm,width=8.0cm]{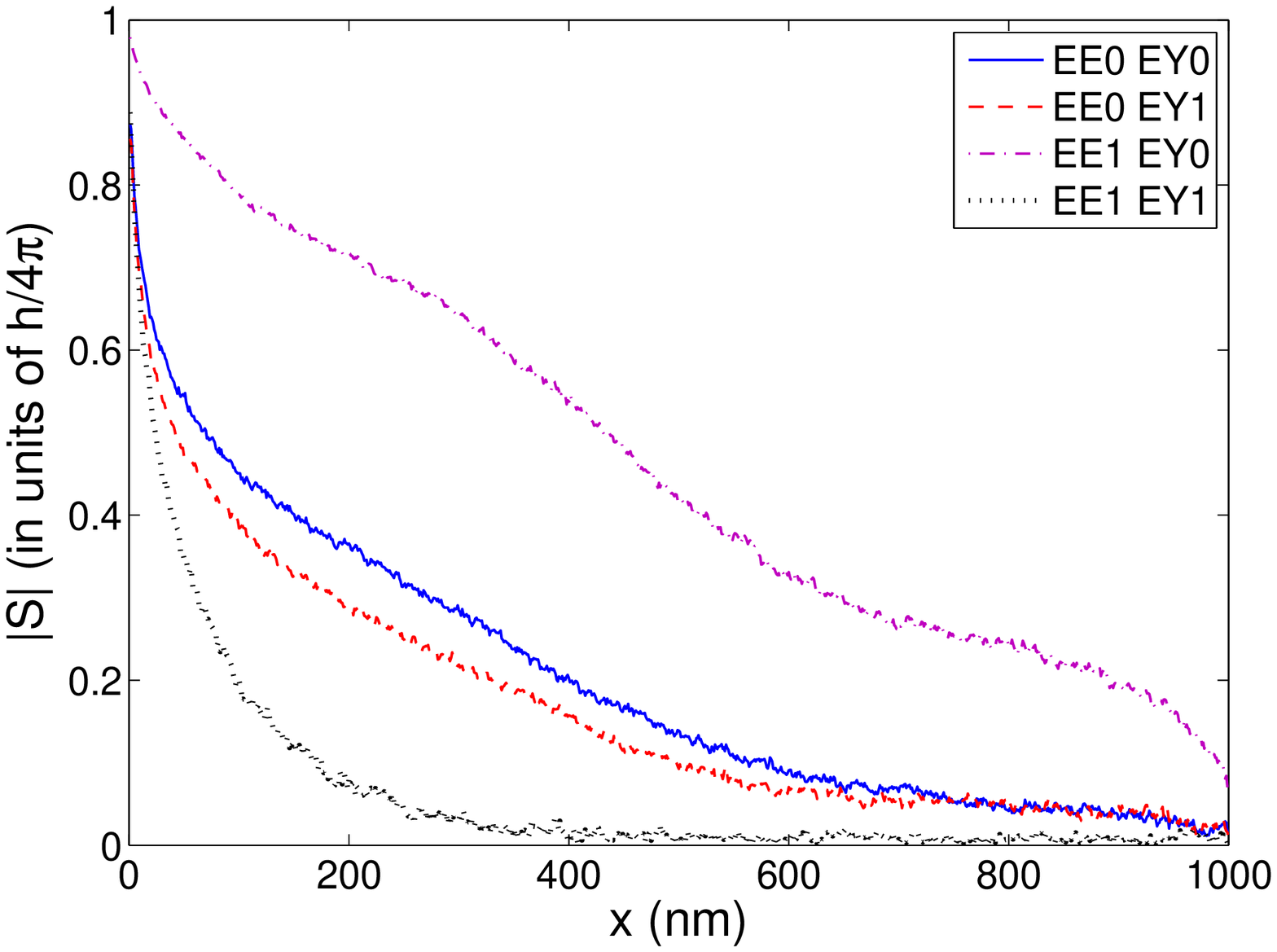}\label{InSb}}
\quad
\subfloat[][]
{\includegraphics[height=6.3cm,width=8.0cm]{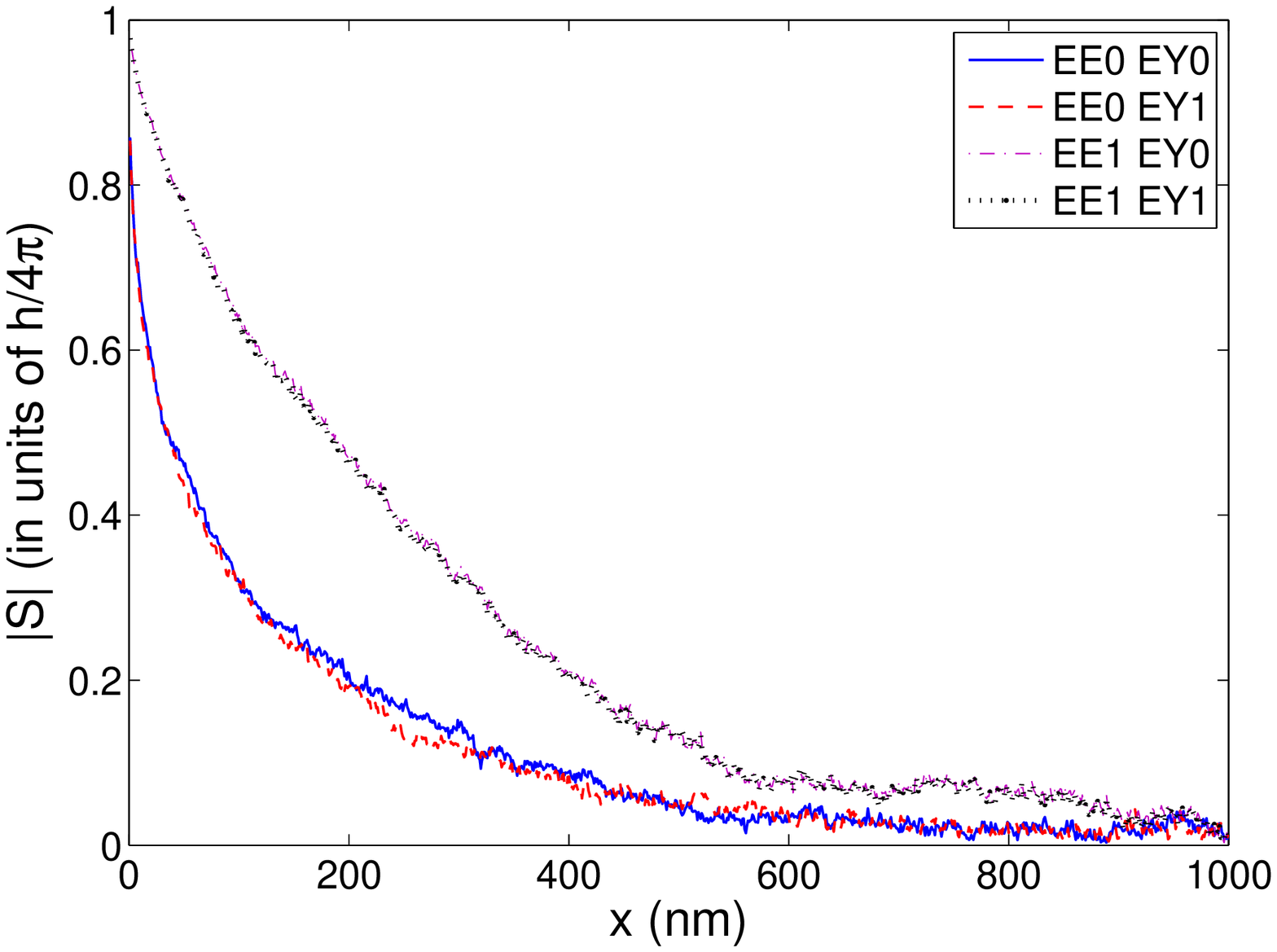}\label{GaAs}}
\caption{$|S|$ along the length of the device for the four permutations possible by considering or not considering electron-electron scattering and Elliott-Yafet spin relaxation mechanism in (a) InSb and (b) GaAs based heterostructures. EY and EE stand for Elliott-Yafet mechanism and electron-electron scattering respectively. The label 1 or 0 means that the corresponding scattering mechanism has been included or neglected, respectively, in the simulations.}
\end{figure}

The simulation results were obtained for the four cases possible by accounting for and not accounting for electron-electron scattering and EY spin relaxation mechanism in the simulation routines. Figure \ref{InSb} depicts these results for the InSb based structure. As compared to the case in which neither of the two phenomena are considered (which we choose to call the basic case), only a minute decline in the spin relaxation length is noticed on inclusion of the EY spin relaxation mechanism. This points to the dominance of DP mechanism in causing decoherence when electron-electron scattering is not considered. On the other hand, when EY mechanism is not considered and electron-electron scattering is taken into account, the simulations predict a much higher spin relaxation length as compared to the basic case. This could have been anticipated based on the results discussed in the preceding subsection. The accurate results, however, are arrived at when both the phenomena are considered, in which case, the spin relaxation length declines appreciably as compared to the basic case establishing EY mechanism as dominant in causing spin decoherence in the InSb system considered. On the other hand, in the GaAs based structure, EY mechanism has been found to play no role whatsoever, and DP mechanism is obviously the dominant spin decoherence mechanism (Fig. \ref{GaAs}). 

We pause to compare our results on spin relaxation length with analogous results on spin relaxation time obtained by Jiang and Wu \cite{wu5}. Jiang and Wu have found dominance of DP mechanism in deciding the spin relaxation time in an initially polarized spatially homogeneous InSb system. One may be tempted to extend this finding to spin relaxation length as well. We must caution that given the special property of electron-electron scattering that it contributes to momentum redistribution without causing momentum relaxation, and the already shown relative weakness of DP mechanism in causing spin decoherence in spin transport \cite{kamra}, the dominance reversal in deciding spin relaxation length can very well be expected. This is especially true as in Jiang and Wu's work, the role of EY mechanism is seen to increase with increasing carrier density, and for a density equivalent to that considered in the present work, one can extrapolate the ratio $\tau_{EY}/\tau_{DP}$ to be of the order of $1$ at $300 K$.

Hence, we find that electron-electron scattering results in the dominance of EY mechanism in high mobility InSb based heterostructure. A possible way to experimentally determine the dominant spin decoherence mechanism is to investigate the spin relaxation length as a function of electron concentration while making sure that the latter is small enough so that the dynamics is confined to the lowest subband. An increase in spin relaxation length with carrier concentration will indicate the dominance of DP mechanism and a decrease in $L_{s}$ will indicate dominance of EY mechanism. It should be noted that although the dominance of EY mechanism has been obtained for InSb, it may be true for narrow gap semiconductors in general. As the bandgap increases, EY mechanism becomes weaker and spin relaxation length increases until DP mechanism begins to dominate. Therefore, wide gap semiconductors may be preferred over their narrow gap counterparts for use in spintronics.

\section{Conclusion}
The importance of electron-electron scattering in spin transport has been demonstrated using Monte Carlo simulation of spin transport in \textrm{III-V} semiconductor heterostructures. The electron-electron scattering is necessary to obtain an accurate distribution in the $k$ space. Its inclusion in the theory predicts an appreciable change in the dominance domains of DP and EY spin relaxation mechanisms. Our simulations predict the dominance of EY spin relaxation mechanism in narrow gap semiconductor based heterostructures. Our work shows that the spin relaxation length is expected to increase with increasing bandgap until DP mechanism begins to dominate. It has already been shown in a space resolved analysis of DP mechanism \cite{kamra} that longer spin relaxation lengths are favoured by smaller $\beta <k_{z}^{2}>/\alpha$. This, combined with the fact that this ratio is smaller for lower bandgap materials in general \cite{bandyopadhyay}, can be used to argue that highest spin relaxation length can be expected in a medium band gap material in which EY relaxation mechanism is weakened. $\mbox{In}_{0.53}\mbox{Ga}_{0.47}\mbox{As}$ might be such a material.

\section*{Acknowledgement}
The authors are grateful to T. K. Ghosh for useful discussions. We also thank M. M. Glazov, E. L. Ivchenko and M. W. Wu for valuable comments. This work was supported by funding from Ministry of Human Resource Development (MHRD), India.


\begin{thebibliography}{40}

\bibitem{fabian}
I. Zutic, J. Fabian, and S. Das Sarma, Rev. Mod. Phys. {\bf 76}, 323 (2004).

\bibitem{bandyopadhyay}
S. Bandyopadhyay and M. Cahay, Introduction to Spintronics (CRC, Boca Raton, 2008).

\bibitem{zutic}
J. Fabian, A. Matos-Abiague, C. Ertler, P. Stano, and I. Zutic,  Acta Physica Slovaca {\bf 57}, 565 (2007).

\bibitem{wu}
M. W. Wu, J. H. Jiang, and M. Q. Weng, Physics Reports {\bf 493}, 61 (2010).

\bibitem{dp}
M. I. D'yakonov and V. I. Perel', Sov. Phys. Solid State {\bf 13}, 3023 (1972).

\bibitem{ey}
J. Elliott, Phys. Rev. {\bf 96}, 266 (1954).

\bibitem{jacoboni}
C. Jacoboni and L. Reggiani, Rev. Mod. Phys. {\bf 55}, 645 (1983).

\bibitem{lugli}
C. Jacoboni and P. Lugli, The Monte Carlo Method for Semiconductor Device Simulation (Springer-Verlag, Wien, 1989).

\bibitem{warmenbol}
P. Warmenbol, F. M. Peeters, J. T. Devreese, G. E. Alberga, and R. G. van Welzenis, Phys. Rev. B {\bf 31}, 5285 (1985).

\bibitem{kocevar}
P. Lugli, C. Jacoboni, L. Reggiani, and P. Kocevar, Appl. Phys. Lett. {\bf 50}, 1251 (1987).

\bibitem{yokoyama}
K. Yokoyama and K. Hess, Phys. Rev. B {\bf 33}, 5595 (1986).

\bibitem{chattopadhyay}
D. Chattopadhyay and A. Bhattacharyya, Phys. Rev. B {\bf 37}, 7105 (1988).

\bibitem{privman}
S. Saikin, M. Shen, M. C. Cheng, and V. Privman, J. Appl. Phys. {\bf 94}, 1769 (2003).

\bibitem{kamra}
A. Kamra, B. Ghosh, and T. K. Ghosh, J. Appl. Phys. {\bf 108}, 054505 (2010).

\bibitem{tierney2}
B. D. Tierney and S. M. Goodnick, Proceedings of the 7th IEEE International Conference on Nanotechnology, August 2 - 5, 2007, Hong Kong.

\bibitem{saikin}
S. Saikin, M. Shen, and M. C. Cheng, IEEE Transactions on Nanotechnology {\bf 3}, 173 (2004).

\bibitem{glazov}
M. M. Glazov and E. L. Ivchenko, JETP Letters {\bf 75}, 403 (2002).

\bibitem{glazov2}
M. M. Glazov and E. L. Ivchenko, JETP {\bf 99}, 1279 (2004).

\bibitem{wu1}
M. Q. Weng and M. W. Wu, J. Appl. Phys. {\bf 93}, 410 (2003).

\bibitem{wu2}
J. L. Cheng and M. W. Wu, J. Appl. Phys. {\bf 101}, 073702 (2007).

\bibitem{wu3}
J. L. Cheng, M. W. Wu, and I. C. da Cunha Lima, Phys. Rev. B {\bf 75}, 205328 (2007).

\bibitem{wu4}
P. Zhang and M. W. Wu, Phys. Rev. B {\bf 79}, 075303 (2009).

\bibitem{songkim}
P. H. Song and K. W. Kim, Phys. Rev. B {\bf 66}, 035207 (2002).

\bibitem{bh}
H. Brooks and C. Herring, Phys. Rev. {\bf 83}, 879 (1951).

\bibitem{mosko}
M. Mosko and A. Moskova, Phys. Rev. B {\bf 44}, 10794 (1991).

\bibitem{ferry}
P. Lugli and D. K. Ferry, Physica B {\bf 117}, 251 (1983).

\bibitem{brunetti}
R. Brunetti, C. Jacoboni, V. Dienys, and A. Matulionis, Physica B {\bf 134}, 369 (1985).

\bibitem{rashba}
Y. A. Bychkov and E. I. Rashba, J. Phys. C {\bf 17}, 6039 (1984).

\bibitem{dresselhaus}
G. Dresselhaus, Phys. Rev. {\bf 100}, 580 (1955).

\bibitem{pikus}
G. E. Pikus and A. N. Titkov, in Optical Orientation, edited by F. Meier and B. P. Zakharchenya ͑(North-Holland, Amsterdam, 1984͒).

\bibitem{tamborenea}
P. I. Tamborenea, M. A. Kuroda, and F. L. Bottesi, Phys. Rev. B {\bf 68}, 245205 (2003).

\bibitem{dattabook}
S. Datta, Electronic Transport in Mesoscopic Systems (Cambridge University Press, New York, 1997).

\bibitem{nitta}
J. Nitta, T. Akazaki, H. Takayanagi, and T. Enoki, Phys. Rev. Lett. {\bf 78}, 1335 (1997).

\bibitem{damocles}
M. V. Fischetti and S. E. Laux, Damocles Theoretical Manual (IBM Corporation, Yorktown Heights, 1995).

\bibitem{gilbertson}
A. M. Gilbertson, W. R. Branford, M. Fearn, L. Buckle, P. D. Buckle, T. Ashley, and L. F. Cohen, Phys. Rev. B {\bf 79}, 235333 (2009).

\bibitem{tierney}
B. D. Tierney, T. E. Day, and S. M. Goodnick, Journal of Physics: Conference Series {\bf 109}, 012034 (2008).

\bibitem{wu5}
J. H. Jiang and M. W. Wu, Phys. Rev. B {\bf 79}, 125206 (2009).


\end{thebibliography}
\end{document}